\documentclass[prb,showpacs,twocolumn,aps,superscriptaddress,a4paper]{revtex4-1}
\usepackage{dcolumn,amssymb,amsmath,amsfonts,graphicx,latexsym,color}

\begin{document}

\long\def\/*#1*/{}

\title{The transport properties of Floquet topological superconductors at the transition from the topological phase to the Anderson localized phase}

\author{Pei Wang}
\affiliation{International Center for Quantum Materials, School of Physics, Peking University, Beijing 100871, China}
\affiliation{Institute of Applied Physics, Zhejiang University of Technology, Hangzhou 310023, China}
\author{Qing-feng Sun}
\affiliation{International Center for Quantum Materials, School of Physics, Peking University, Beijing 100871, China}
\affiliation{Collaborative Innovation Center of Quantum Matter, Beijing 100871, China}
\author{X. C. Xie}
\affiliation{International Center for Quantum Materials, School of Physics, Peking University, Beijing 100871, China}
\affiliation{Collaborative Innovation Center of Quantum Matter, Beijing 100871, China}
\date{\today}

\begin{abstract}
The Floquet topological superconducting state is a nonequilibrium time-periodic state hosting Majorana fermions. We study its transport properties by using the Kitaev model with time-periodic incommensurate potentials, which experiences phase transition from the Floquet topological superconducting phase to the Anderson localized phase with increasing driving strength. We study both the real time dynamics of the current and the non-analytic behavior of the tunneling conductance at the transition. Especially, we find that the tunneling conductance changes continuously at the transition, being a finite value in the presence of Floquet Majorana fermions, but dropping to zero as the Majorana fermions vanish. For a special choice of parameters, the Majorana fermions revive at larger driving strength, accompanied by the revival of conductances.
\end{abstract}

\maketitle

\section{Introduction}

Whether Majorana fermions exist in real world keeps an unsolved problem since they were first proposed~\cite{majorana} more than seventy years ago. Recently the many-body systems become the important candidates for hosting Majorana fermions~\cite{beenakker} which may exist there as collective excitations. Several solid state and cold atomic systems have been proposed~\cite{moore91,read00,rice95,sarma06,gurarie05,tewari07,fu08}, e.g., the topological superconductors~\cite{fu08}, while the detection of Majorana fermions keeps a challenge. A promising way is by the transport properties of Majorana edge states, i.e., a quantized conductance at zero temperature at the junction of the system and a normal metal~\cite{law09}. Recent experiments on quantum wires~\cite{mourik12,deng12,das12} reported the observation of zero bias peaks in tunneling conductances at finite temperatures, but did not provide the exclusive evidence of Majorana fermions, because the impurities in the materials may cause similar peaks~\cite{
pikulin12,liu12}.

The Floquet topological theory provides a new approach of making Majorana fermions dynamically. Kitagawa {\it et al.}~\cite{kitagawa10} extended the definition of topological order from equilibrium states to periodically driven states by Floquet theory. Based on this definition, some authors~\cite{jiang,liu12b,liu13,reynoso13} predicted the Floquet topological superconductor (superfluid), which host Majorana fermions at their edges. And these Floquet Majorana fermions have the same non-Abelian statistics as their equilibrium counterparts~\cite{liu13}.

Concerning the Floquet Majorana fermions, much less is known about their detection. A quantized conductance signals the Majorana fermions in equilibrium topological superconductors. However, the Floquet topological superconductor is in a time-periodic nonequilibrium state, in which a (quasi)particle may change its energy by absorbing or releasing photons. The lack of a conserved quasiparticle energy modifies the transport process. The resonant tunneling is accompanied by the photon-assisted tunneling, so that the linear conductance is not quantized any more even at zero temperature. Kundu and Seradjeh~\cite{kundu13} derived a sum rule for linear conductances, but it is difficult to measure in experiments. Further studies on the transport properties of Floquet topological superconductors are still necessary.

On the other hand, incommensurate potential was recently employed for studying the effect of disorder on the transport properties of topological superconductors due to its good controllability. The incommensurate potential has been experimentally engineered in cold atomic systems~\cite{Roati}, and its generation in solid state systems was also under discussion~\cite{gang12}. Studies on the Kitaev's p-wave model with incommensurate potentials~\cite{degottardi,cai13} predicted a topological phase transition from the topological superconducting phase to the Anderson localized phase as the disorder strength increases. And the tunneling conductance shows a jump at the critical point~\cite{pei13b}. This jump belongs to the non-analytic behaviors at an equilibrium phase transition. Similarly, we expect the non-analytic behavior of conductances in a time-periodically driven system, when it experiences a nonequilibrium phase transition by tuning the driving field~\cite{liu13}. This non-analytic behavior provides a 
possible way of detecting Majorana fermions since the direct observation of a quantized conductance is still impossible. It is then worth of studying the evolution of conductances with changing driving field.

Therefore, we study the tunneling conductance of the Kitaev's p-wave model with time-periodic incommensurate potentials in this paper. By choosing appropriate parameters, we find that this system is in the Floquet topological superconducting phase in the weak driving regime, but experiences phase transition to a Floquet topologically trivial phase as the driving strength increases. The conductance of Floquet topological superconductors drops to zero at the transition, being a continuous but non-smooth function. And the zero conductance is accompanied by the vanishing of Majorana edge states. Our findings contribute to the problem of detecting Floquet Majorana fermions by showing the details of the non-analytic behavior of conductances at the transition.

The transport properties of a driven quantum system are usually studied by the Floquet Green's function method~\cite{kundu13}. While considering that the current is time-dependent in our system, we develop the recently introduced numerical operator method~\cite{pei12} for our calculations. Our method gives not only averaged conductances, but also the real time dynamics of currents in one period. The latter is absent in previous studies. We then provide a complete description of the transport properties of Floquet topological superconductors including their time-dependent characteristics. 

The rest of paper is organized as follows. In Sec.~\ref{sect:model}, we discuss the topological order of the Kitaev's model with time-periodic incommensurate potentials by calculating the Floquet spectrum. In Sec.~\ref{sect:method}, we introduce the transport setup and our method for calculating the tunneling currents and conductances. The results of tunneling currents are discussed in Sec.~\ref{sect:current}. And the linear conductances are discussed in Sec.~\ref{sect:conductance}. Sec.~\ref{sect:conclusion} is a short summary.

\section{Topological order of the Kitaev's model with time-periodic incommensurate potentials}
\label{sect:model}

We start from the Kitaev model on an incommensurate lattice~\cite{cai13}. Its Hamiltonian is written as
\begin{equation}\label{hamiltoniansc}
\begin{split}
  \hat H_0 = & -\sum_{i=1}^{L-1} g_s (\hat c^\dag_i \hat c_{i+1} + h.c.) + \Delta \sum_{i=1}^{L-1} (\hat c^\dag_i \hat c^\dag_{i+1} + h.c.) \\ & + \sum_{i=1}^{L} \mu_i \hat c^\dag_i \hat c_i,
\end{split}
\end{equation}
where $L$ denotes the chain length, $g_s$ the hopping and $\Delta$ the superconducting pairing. The incommensurate potential is generally expressed as $\mu_i=V_{d} \cos(2\pi i\alpha)$ with $\alpha=(\sqrt{5}-1)/2$ and $V_d$ denoting the disorder strength. The model~(\ref{hamiltoniansc}) describes a p-wave topological superconductor with two Majorana bound states at the ends in the weak disorder regime $V_d\leq 2g_s +2\Delta$. The system experiences a topological phase transition at $V_d =2g_s +2\Delta$ and is driven into the Anderson localized phase in the presence of strong disorder~\cite{cai13}.

We drive the system out of equilibrium by a homogeneous periodic potential, which can be generated by side-gates in quantum wires or by laser beams in optical lattices. The driving Hamiltonian is
\begin{equation}
\hat H_I(t)=m\sin(\Omega t)\sum_{i=1}^{L} \hat c^\dag_i \hat c_i,
\end{equation}
where $m$ denotes the driving strength and $\Omega$ the driving frequency. The total Hamiltonian of the driven system is 
\begin{equation}\label{hamiltoniandriving}
\hat H(t) = \hat H_0 + \hat H_I(t).
\end{equation}

\begin{figure}
\includegraphics[width=1\linewidth]{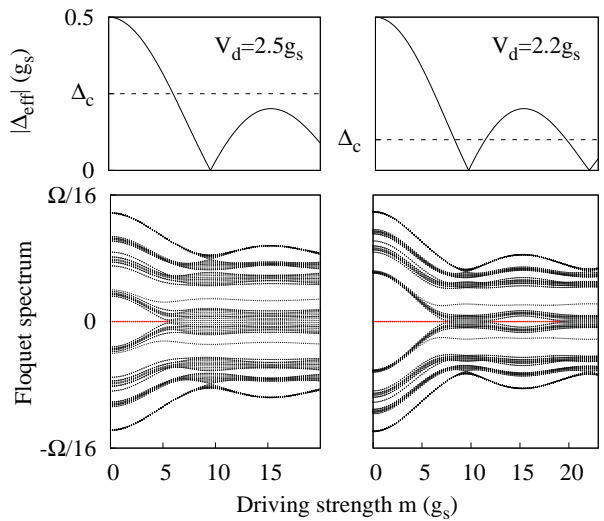}
\caption{The top panels show the effective superconducting pairing $|\Delta_{eff}|$ as a function of $m$, while the critical pairing $\Delta_c$ is marked by dashed lines. The bottom panels show the corresponding Floquet spectrum of $\hat H(t)$ respectively. The zero energy modes are marked in red color. We set $\Omega=8g_s$, $\Delta=0.5g_s$ and $L=200$. The left top and bottom panels are at $V_d=2.5g_s$, while the right ones are at $V_d=2.2g_s$.}\label{fig:floquetspec}
\end{figure}
We first qualitatively discuss the topological order of this driven system. Following the approximation by Liu {\it et al.}~\cite{liu13}, $\hat H(t)$ can be transformed into a stationary effective Hamiltonian
\begin{equation}
\begin{split}
\hat H_{eff} = & -\sum_{i=1}^{L-1} g_s (\hat c^\dag_i \hat c_{i+1} + h.c.) + \Delta_{eff} \sum_{i=1}^{L-1} (\hat c^\dag_i \hat c^\dag_{i+1} + h.c.) \\ & + \sum_{i=1}^{L} \mu_i \hat c^\dag_i \hat c_i.
\end{split}
\end{equation}
The effective Hamiltonian $\hat H_{eff}$ has the same form as $\hat H_0$ except that $\Delta$ is replaced by the effective superconducting pairing $\Delta_{eff}=\Delta J_0 (\frac{2m}{\Omega} )$ where $J_0$ denotes the Bessel function of zeroth order. Because $|J_0 (\frac{2m}{\Omega})|\leq 1$ and then $|\Delta_{eff}|\leq \Delta$, the sole effect of the periodic potential is to suppress the superconducting pairing. Since a topological phase transition happens at $V_d=2g_s+2|\Delta_{eff}|$ in the Hamiltonian $\hat H_{eff}$, we deduce that the driven system experiences phase transition at
\begin{equation}\label{besselcondition}
 \begin{split}
 |\Delta_{eff}| = & \Delta_c \\
 := & (V_d -2 g_s)/2,
 \end{split}
\end{equation}
once the disorder strength satisfies $(2\Delta+2g_s)>V_d >2g_s$. The system is in the Anderson localized phase for $|\Delta_{eff}| < \Delta_c$ and in the topological superconducting phase for $|\Delta_{eff}| > \Delta_c$.

A particularly interesting result of Eq.~(\ref{besselcondition}) is got from the oscillation of the Bessel function in $\Delta_{eff}$. If $\Delta_c/\Delta$ is sufficiently small, the system is in the topological superconducting phase at $m=0$. By increasing the driving strength, we get the topologically trivial phase. But when further increasing $m$, we resurrect the topological superconducting phase (see the top right panel of Fig.~\ref{fig:floquetspec}). The oscillation of $\Delta_{eff}$ results in the alternate disappearance and appearance of Majorana fermions. From the shape of the function $|\Delta_{eff}|$, it is easy to see that the existence of multiple transitions as $m$ increasing is only possible if $V_d$ is close to $2g_s$.

The qualitative analysis requires the driving frequency being much larger than the other energy scales, i.e., $\Omega\gg g_s,\Delta$. So we take $\Omega=8g_s$ throughout this paper. At the same time, for a serious discussion of the topological order, we calculate the Floquet spectrum of $\hat H(t)$ under open boundary conditions. The topological order is precisely determined by whether there are zero modes.

The Hamiltonian satisfies $\hat H(t)=\hat H(t+T)$ where $T=2\pi/\Omega$ is the period. We re-express $\hat H(t)$ in a matrix form as
\begin{eqnarray}
 \hat H(t) = \frac{1}{2} \left( \hat c^\dag_1 \cdots \hat c^\dag_L \hat c_1 \cdots \hat c_L \right) \check H(t) \left( \begin{array}{c} \hat c_1 \\ \vdots \\ \hat c_L \\ \hat c_1^\dag \\ \vdots \\ \hat c_L^\dag \end{array} \right),
\end{eqnarray}
and suppose the quasiparticle state to be
\begin{equation}
|\psi(t)\rangle = \sum_j( \psi_{u j}(t) \hat c^\dag_j + \psi_{v j}(t) \hat c_j) |0\rangle,
\end{equation}
where $|0\rangle$ denotes the quasiparticle ground state. The Schr\"{o}dinger equation in this picture is written as
\begin{equation}\label{eigenequationphi}
 \check H(t) \psi(t) = i\partial_t \psi(t),
\end{equation}
where $\psi(t) = (\psi_{u1}, \cdots, \psi_{uL},\psi_{v1},\cdots,\psi_{vL})^T$. According to the Floquet theorem, we have $\psi(t)= e^{-i\epsilon t} \phi(t)$, where $\phi(t)=\phi(t+T)$ is periodic and satisfies the eigen equation
\begin{equation}\label{floquetmatrix}
 (\check H(t) -i \partial_t)\phi(t)= \epsilon \phi(t). 
\end{equation}
The Floquet spectrum is the set of quasienergies $\epsilon$.

We divide the Hamiltonian matrix into the time-independent part and the driving part: $\check H(t)= \check H_0 + \check H_I(t)$, where $\check H_I(t)=m\sin(\Omega t) \check \sigma_z \otimes \check{I}$ with $\check{\sigma}_z$ the Pauli matrix and $\check{I}$ the identity matrix of dimension $L$. It is favorable to decompose the wave function $\phi(t)$ in the eigen basis of $\check{H}_0$. We suppose the eigenvectors of $\check H_0$ to be $\phi^{(0)}_\alpha$ with the corresponding eigenvalue $e_\alpha$. Noting that $\phi(t)$ is periodic, we express the wave function as
\begin{equation}\label{phidecompositioin}
 \phi (t) = \sum_{\alpha,n=-\infty}^\infty C_{n,\alpha} e^{-in\Omega t} \phi^{(0)}_\alpha,
\end{equation}
where $n$ is an integer and the factor $e^{-in\Omega t}$ comes from the periodicity of $\phi(t)$. Substituting Eq.~(\ref{phidecompositioin}) into Eq.~(\ref{floquetmatrix}), we obtain the eigen equation of the coefficient array $C_{n,\alpha}$:
\begin{equation}
\sum_{\alpha,n=-\infty}^\infty \check{h}_{n'\beta,n\alpha} C_{n\alpha} =\epsilon C_{n'\beta},
\end{equation}
where the elements of $\check{h}$ are expressed as
\begin{equation}
\begin{split}
\check{h}_{n'\beta,n\alpha} = & ( e_{\alpha}-n\Omega)\delta_{n,n'}\delta_{\alpha,\beta} \\ & + \frac{m}{2i} (\delta_{n',n-1}-\delta_{n',n+1}) \phi_{\beta}^{(0)\dag} \check{\sigma}_z \otimes \check I \phi_{\alpha}^{(0)} .
\end{split}
\end{equation}
We numerically diagonalize the matrix $\check{h}$ to get the quasienergies $\epsilon$. These quasienergies are packed into a series of Floquet bands of width $\Omega$. Different Floquet bands repeat each other except an overall energy shift. The Floquet spectrum usually means the band within the interval $[-\Omega/2,\Omega/2]$.

The bottom panels of Fig.~\ref{fig:floquetspec} plots the Floquet spectrum of $\hat H(t)$ as the driving strength $m$ varies. While the top panels show $|\Delta_{eff}|$ as a function of $m$ together with the critical superconducting pairing $\Delta_c$. The Floquet spectrum roughly agrees with our judgement of topological order by comparing $|\Delta_{eff}|$ and $\Delta_c$. At small $m$, the Floquet spectrum has an energy gap and zero modes, indicating that the system is in the Floquet topological superconducting phase. As $m$ increases, the gap closes at the critical point of the Floquet topological phase transition. Beyond this critical point, the spectrum shows the feature of the Anderson localized phase, i.e., there is no energy gap while the quasienergies are very close to but not strictly zero. The loss of zero modes indicates a Floquet topologically trivial phase. But one should notice that the system is in a time-dependent nonequilibrium state and then is not in the true Anderson localized 
phase. 

The alternate appearance and disappearance of the topological superconducting phase is observed at $V_d=2.2g_s$ in the Floquet spectrum (see the right bottom panel of Fig.~\ref{fig:floquetspec}). As $m$ increases, the energy gap alternately closes and opens. And the opening of the gap is always accompanied by the appearance of zero modes. This result coincides with the above analysis of an oscillating $|\Delta_{eff}|$. 

The critical points of the transition can be quantitatively decided from the Floquet spectrum. At $V_d=2.5g_s$, there is a single critical point at $m \approx 5.4g_s$. At $V_d=2.2g_s$, there exist multiple critical points at $m \approx 7.6g_s, 12.0g_s, 19.0g_s, \ldots$.

\section{Transport setup and numerical operator method}
\label{sect:method}

We study the transport properties of Floquet Majorana fermions by the tunneling currents and conductances between normal leads and the driven Kitaev chain. In our setup, the Kitaev chain is located between two semi-infinite leads, coupled at the end sites. The Hamiltonians of the left and right leads are expressed as
\begin{equation}
 \hat H_L = -g_l \sum_{i=-\infty}^{-1} (\hat c^\dag_i \hat c_{i+1} + h.c.)
\end{equation}
and
\begin{equation}
\hat H_R = - g_l \sum_{i=L+1}^\infty  (\hat c^\dag_i \hat c_{i+1} + h.c.)
\end{equation}
respectively. The coupling Hamiltonian between the leads and the Kitaev chain is expressed as
\begin{equation}
 \hat H_V = g_c (\hat c^\dag_{0} \hat c_{1} + \hat c^\dag_{L} \hat c_{L+1} + h.c.).
\end{equation}
Then the total Hamiltonian of the transport setup is
\begin{equation}
 \hat H_t(t) = \hat H (t) + \hat H_L + \hat H_R + \hat H_V,
\end{equation}
where $\hat H(t)$ describes the driven Kitaev chain and has been given in Eq.~(\ref{hamiltoniandriving}).

The two leads are in different chemical potentials with $\mu_L=V$ and $\mu_R=0$ respectively, where $V$ is just the voltage bias across the transport setup. The tunneling current from the left lead to the Kitaev chain is written as
\begin{equation}\label{leftcdef}
\begin{split}
 I_L (t) = -2 g_c \textbf{Im} \langle \hat c^\dag_{0}(t) \hat c_1(t) \rangle ,
\end{split}
\end{equation}
and that from the Kitaev chain to the right lead is
\begin{equation}\label{rightcdef}
\begin{split}
 I_R (t) = -2 g_c \textbf{Im} \langle \hat c^\dag_{L}(t) \hat c_{L+1}(t) \rangle.
\end{split}
\end{equation}
The averaged current is defined as $I(t)=(I_L(t)+I_R(t))/2$.

The key point of the numerical operator method is calculating $\hat c^\dag_0(t)$, $\hat c_1(t)$, $\hat c^\dag_{L}(t)$ and $\hat c_{L+1}(t)$ in the Heisenberg picture, i.e., expressing them by the original field operators $\hat c_j$ and $\hat c^\dag_{j'}$. We set $t_0=0$ the initial time, and divide $t$ into $N$ small time intervals of length $\tau$. According to the definition, the field operator in the Heisenberg picture is expressed as
\begin{equation}
\begin{split}
 \hat c^\dag_0 (t)= & e^{i\hat H_t(t_0) \tau} \cdots e^{i\hat H_t \left(t_{N-1} \right) \tau} \hat c^\dag_0 \\ & \times e^{-i\hat H_t \left(t_{N-1} \right) \tau} \cdots e^{-i\hat H_t(t_0) \tau},
\end{split}
 \end{equation}
where $t_{j}=j\tau$. Calculating $\hat c^\dag_0(t)$ is finished in $N$ steps. In the {\it n}-th step, we undress the pair of operators $e^{i\hat H_t \left(t_{N-n} \right) \tau} $ and $e^{-i\hat H_t \left(t_{N-n} \right) \tau} $ from the central operator $\hat c^\dag_0$. For example, the first step is to get
\begin{equation}
\begin{split}
e&^{i\hat H_t \left(t_{N-1} \right) \tau} \hat c^\dag_0  e^{-i\hat H_t \left(t_{N-1} \right) \tau} \\ = & \hat c^\dag_0 + i\tau [\hat H_t \left(t_{N-1} \right),  \hat c^\dag_0 ] \\ & +\frac{(i\tau)^2}{2} [\hat H_t \left(t_{N-1} \right), [\hat H_t \left(t_{N-1} \right),  \hat c^\dag_0 ]] + O(\tau^3).
\end{split}
\end{equation}
After each step, we get a linear combination of original field operators. The length of this expression increases linearly with steps. A truncation scheme is employed to keep the expression in a finite length (see more details in the Ref.~\cite{pei13b}).

It is worth of mentioning a trick in the numerical calculation. Considering that $\hat H_t(t)$ is periodic, we always choose $\tau$ so that the period $T$ is an integer times of $\tau$, and $\tau$ is decided adaptively to make the discretization error negligible. This trick is critical for obtaining the accurate result at a large $t$.

After getting the expressions of $\hat c^\dag_0(t)$ {\it et al.}, the calculation of $I_L(t)$ and $I_R(t)$ is straight forward. We substitute the expressions into Eq.~(\ref{leftcdef}) and (\ref{rightcdef}) and calculate their expectation values to the initial state. It is necessary to make clear the initial state here. At $t_0=0$, we suppose that the leads and the Kitaev chain are decoupled to each other. The leads are in the ground states of $\hat H_L$ and $\hat H_R$ with the chemical potentials $\mu_L$ and $\mu_R$ respectively. While the Kitaev chain is in the ground state of $\hat H_0$. The expressions of expectation values can be found in the Ref.~\cite{pei13b}.

This choice of the initial state indicates that $t$ must be sufficiently large for the initial correlation being forgotten and the periodic currents being obtained. In practice, we calculate the currents at $t=200/g_s\sim 300/g_s$, i.e., several hundreds of periods from the initial time. We compare the currents at different periods to make sure that the initial condition is irrelevant. It is worth of mentioning that our method guarantees the results being accurate at large times, because we conquer the finite size problem by taking infinite leads~\cite{pei13b}.

\begin{figure}
\includegraphics[width=1\linewidth]{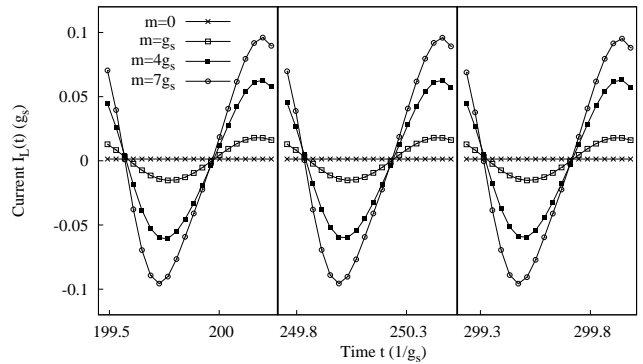}
\caption{The time-dependent currents $I_L(t)$ within three complete periods at $t=200/g_s,250/g_s$ and $300/g_s$ are plotted in the left, center and right panels respectively. Different type of points refer to different driving strength $m$. The solid lines serve as guides for the eyes. We set $\Omega=8g_s$ (corresponding to a period $T\approx 0.785/g_s $), $\Delta=0.5g_s$ and $V_d=2.5g_s$. The chain length is set to be sufficiently large as $L=1000$. The parameters of the leads and the coupling is set to $g_l=10g_s$ and $g_c=g_s$. And the voltage bias is chosen to be small as $V=0.005g_s$.}\label{fig:currentperiod}
\end{figure}
Compared with the currents, the linear conductances are more relevant in experiments and are calculated separately in this paper. Corresponding to $I_L(t)$, $I_R(t)$ and $I(t)$, the linear conductances at the zero bias are defined as $G_L(t)= \frac{\partial I_L(t)}{\partial V}|_{V=0}$, $G_R(t)=\frac{\partial I_R(t)}{\partial V}|_{V=0}$ and $G(t)=\frac{\partial I(t)}{\partial V}|_{V=0}$ respectively. Since the field operators $\hat c^\dag_0(t)$ {\it et al.} in the expressions of currents are independent to $V$, the partial differential operator acts only upon the expectation values. We can then express the linear conductances as
\begin{equation}\label{lrGdef}
\begin{split}
& G_L (t) = -2 g_c \textbf{Im} \langle \hat c^\dag_{0}(t) \hat c_1(t) \rangle_G , \\
& G_R (t) = -2 g_c \textbf{Im} \langle \hat c^\dag_{L}(t) \hat c_{L+1}(t) \rangle_G,
\end{split}
\end{equation}
where $\langle \cdot \rangle_G = \frac{\partial}{\partial V}\langle \cdot \rangle|_{V=0}$, and $G(t)=(G_L(t)+G_R(t))/2$. Because the expressions of the field operators $\hat c^\dag_0(t)$ {\it et al.} have been obtained in calculating the currents, the left task is to calculate the expectation values $\langle \cdot \rangle_G$. Considering that the initial states of the Kitaev chain and the right lead are independent to $V$, we immediately get
\begin{equation}
 \langle \hat c^\dag_i \hat c_{j} \rangle_G =  \langle \hat c^\dag_i \hat c^\dag_{j} \rangle_G = 0,
\end{equation}
as $i,j \geq 1$. As $i,j\leq 0$, we notice that the left lead is in the ground state of a free Fermi gas at $\mu_L=V$ and then have
\begin{equation}
 \langle \hat c^\dag_i \hat c_j \rangle_G = \frac{\cos \left( (i-j)\displaystyle\frac{\pi}{2} \right)}{2g_l \pi}.
\end{equation}

\section{Tunneling currents}
\label{sect:current}

The evolution of currents is shown in Fig.~\ref{fig:currentperiod}. For comparison, we plot the currents in three complete periods at different times, i.e., $t=200/g_s,250/g_s$ and $300/g_s$. Notice that the period $T=2\pi/\Omega$ is much less than $t$. The oscillation of currents is obvious within one period for $m>0$. While the difference of current-time curves, shown in the left, center and right panels respectively, is negligible at different time. Considering that the time difference between two nearby panels is over $50$ times of periods, we conclude that $t=200/g_s$ is large enough for a good simulation of the periodic current in the limit $t\to \infty$.

In Fig.~\ref{fig:currentperiod}, we see the significant oscillation of currents as the driving field is present, even if the driving strength $m$ is in the localized phase (see the curve at $m=7g_s$). The amplitude of oscillations increases as $m$ increasing for $m\leq 7g_s$. Moreover, the shape of the current-time curves is similar to a trigonometric function, but the distortion is also clear.

\begin{figure}
\includegraphics[width=1\linewidth]{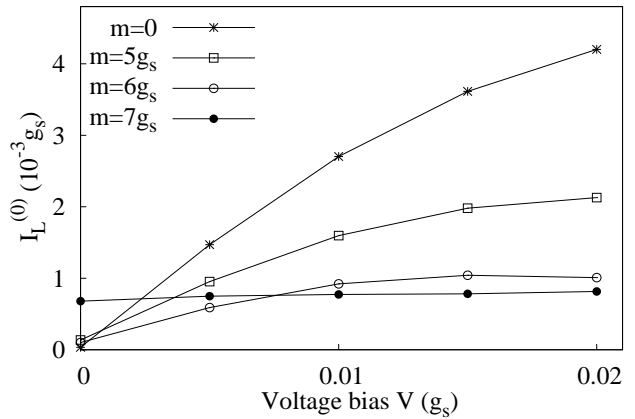}
\caption{The time averaged current in one period as a function of voltage bias at different $m$. Here we set $\Omega=8g_s$, $\Delta=0.5g_s$, $V_d=2.2g_s$, $g_l=10g_s$, $g_c=g_s$ and $L=1000$.}\label{fig:ivcurve}
\end{figure}
For further discussions, we define the time-averaged current $I^{(0)}_L$ as
\begin{equation}
\begin{split}
I^{(0)}_L = \frac{1}{T} \int_T dt I_L(t) ,
\end{split}
\end{equation}
where $\int_T$ denotes the integral within one period. The function $I^{(0)}_L(V)$ at different driving strength is plotted in Fig.~\ref{fig:ivcurve}. The slopes of $I^{(0)}_L$-$V$ curves at $V=0$, which are in fact the linear conductances, gradually decreases as $m$ increases. And the bend of $I^{(0)}_L$-$V$ curves within the range of $V$ is clear, indicating that $V=0.02g_s$ is beyond the linear response regime. Since in the topological superconducting states the linear response regime is decided by the broadening of Majorana levels, we make a conclusion that the level width of the Floquet Majorana states is smaller than $0.02g_s$. The estimation of the level width helps to decide the temperature limit when detecting the Majorana fermions 
in experiments. We also see that the averaged current at the zero bias is not zero in the presence of a driving field, indicating that a net current is driven by the time-periodic potentials even if there is no voltage bias. This is the pumping effect frequently observed in driven systems. 

\section{Differential conductances}
\label{sect:conductance}

\begin{figure}
\includegraphics[width=1\linewidth]{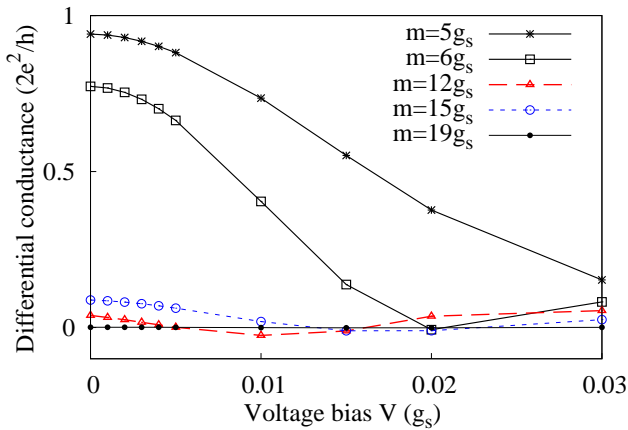}
\caption{(Color online) The differential conductance as a function of voltage bias across the junction at different $m$. We set $V_d=2.2g_s$ and $g_l=3g_s$ in this figure. The lines serve as guides for the eyes.}\label{fig:diffcondtoV}
\end{figure}

The differential conductance is the slope of the current-voltage curve. We set the length of the central superconducting chain to be sufficiently large, much larger than the correlation length of Cooper pairs and that of Majorana bound states at two ends. The right current $I_R(t)$ is then independent to the bias $V$ at the left junction. We immediately get $G_R(t)\equiv 0$ and $G(t)=G_L(t)/2$. We will focus on $G_L$ next. Noting that $I_L(t)\neq I_R(t)$ in our transport setup does not contradict the conservation law of fermions, because there is in fact a third contact to the superconductor in experiments~\cite{mourik12} for grounding it.

We define the time-averaged differential conductance as
\begin{equation}
 G^{(0)}_L = \frac{1}{T} \int_T dt G_L(t).
\end{equation}
This definition coincides with the definition of conductances in previous studies~\cite{kundu13}, in which the conductances are usually calculated by the Green's function method.

The differential conductance as a function of voltage bias is displayed in Fig.~\ref{fig:diffcondtoV}. We clearly see the zero bias peak (a typical feature of Majorana fermions) as the driving strength is small (see the black lines titled $m=5g_s$ and $m=6g_s$). With $m$ increasing the zero bias peak is gradually suppressed and its width shrinks. At $m=12g_s$ (the red line) which has been in the localized phase, the zero bias peak vanishes. However, as $m$ increases further to $m=15g_s$ (the blue line), the zero bias peak reappears even if its height is small, signaling the revival of Majorana fermions. Finally, as $m$ increases to $m=19g_s$, the zero bias peak disappears again. Note that our choice of disorder strength here ($V_d=2.2g_s$) supports the alternate disappearance and reappearance of Majorana bound states.

\begin{figure}
\includegraphics[width=1\linewidth]{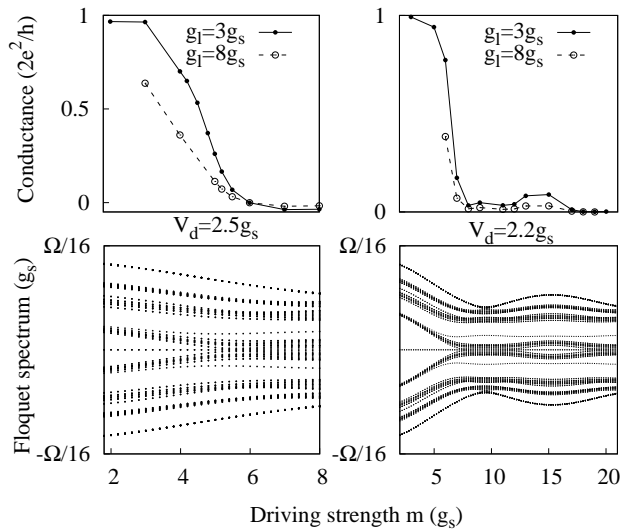}
\caption{The top panels show the tunneling conductances $G^{(0)}_L$ as a function of the driving strength. While the bottom panels show the corresponding Floquet spectrum. The left panels are at $V_d=2.5g_s$, while the right ones are at $V_d=2.2g_s$. Different types of lines represent the conductances at different $g_l$. The other parameters are as same as those in Fig.~\ref{fig:currentperiod}.}\label{fig:conductance}
\end{figure}
Next we study the differential conductance at zero bias, i.e., the linear conductance. We plot $G^{(0)}_L(V=0)$ as a function of the driving strength $m$ in Fig.~\ref{fig:conductance} (the top panels), together with the corresponding Floquet spectrum (the bottom panels). The conductance changes continuously as $m$ increases. No jump is observed in the conductance in spite of the abrupt appearance or disappearance of Majorana bound states (zero modes in the Floquet spectrum) at the Floquet phase transition. This behavior of conductance is well distinguished from that in time-independent systems, e.g., in the Kitaev chain on an incommensurate lattice without the driving, where the conductance jumps from a quantized value to approximately zero as the Majorana bound states vanish at the critical point from the topological superconducting phase to the trivial phase. 

The conductance goes toward $2e^2/h$ as $m\to 0$, coinciding with previous results in time-independent systems. At $V_d=2.5g_s$, the conductance (the left top panel of Fig.~\ref{fig:conductance}) drops as $m$ increases, reaching zero at some point. Comparing the conductance with the Floquet spectrum (the left bottom panel), we find that the conductance drops to zero approximately at the critical point of the Floquet topological phase transition. The conductance being zero beyond the critical $m$ is what we expect, since we have known from the effective Hamiltonian $\hat H_{eff}$ and the Floquet spectrum that the system is in a nonequilibrium state similar to the Anderson localized state. Considering that the conductance function must be non-analytic at the point where it drops to zero and the non-analytic behavior must happen at the phase transition, we conclude that the conductance drops to zero exactly at the critical $m$, even if the errors in the numerical calculation prevent us from obtaining the 
precise $m$ at which $G^{(0)}_L$ reaches zero.

To verify this conclusion, we calculate the conductances at different $g_l$, i.e., the hopping between two neighbor sites in the leads. We see that the tunneling conductances of a driven system depend on the hopping of leads. However, the conductances at different $g_l$ drop to zero at the same $m$ within the numerical errors. This result shows the universality of the behavior that the conductance drops to zero at the transition from the Floquet topological state to the localized state.

The above finding indicates an interesting consequence as $V_d$ is close to $2g_s$. At $V_d=2.2g_s$, as we already know, the Floquet topological superconducting phase alternately disappears and appears as $m$ increases. Correspondingly, the conductance should alternately vanish and reappear. We do observe the revival of conductances at large $m$ in the right top panel of Fig.~\ref{fig:conductance}. In the case of $g_l=8g_s$, the conductance vanishes approximately at $m=7.6g_s$, but then revives approximately at $m=12g_s$. In the case of $g_l=3g_s$, the revival of conductances is much clearer. The deviation of conductances from zero in the region $m\in[7.6g_s,12g_s]$ is attributed to the numerical errors caused by the finite-time effect, which is more important for a smaller $g_l$ because the time for building the periodic current increases with the hopping of leads decreasing. In the range of end time that we can reach, the conductances gradually drop towards zero.

The Floquet topological superconducting state hosts Majorana fermions, but the localized state does not. Our findings then show that the Floquet Majorana fermions do not lead to a quantized conductance. But they are always accompanied by a nonzero conductance, and the vanish of Floquet Majorana fermions is accompanied by the vanish of conductances.

\section{Conclusions}
\label{sect:conclusion}

In summary, the transport properties of a Floquet topological superconducting state is very different from that of a static topological superconducting state. The periodic current has a strong oscillation in course of time, and the amplitude of oscillation may even be larger than the average of current. The average of current is nonzero even at the zero voltage bias due to the pumping effect. And the linear conductance is not quantized, but changes continuously as the driving strength varies. The conductance has no jump at the transition to a topologically trivial phase in which the Majorana bound states vanish. But there are still features in the tunneling conductances which distinguish the Floquet topological phase with Majorana fermions from the Floquet topologically trivial phase without Majorana fermions. These features are shown in the Kitaev model on an incommensurate lattice with time-periodic onsite potentials, which experiences a phase transition from the Floquet topological superconducting phase 
to the Anderson localized phase as the driving strength increases. The 
conductance function is not smooth at the critical point. It is nonzero in the presence of Floquet Majorana fermions, but drops to zero at the transition as the Majorana fermions vanish. For a special choice of parameters, the Floquet Majorana fermions will revive at larger driving strength. Correspondingly, we observe the revival of conductances.

Finally, we would like to mention that the transport properties described in this paper can be measured in experiments. Some proposals have been given for realizing the Kitaev model in solid state or cold atomic systems. The superconducting pairing can be induced by a RF field in cold atomic systems~\cite{jiang} or by the proximity effect in solid state systems~\cite{fulga12}. The periodic incommensurate potentials can be experimentally engineered with ultracold atoms loaded in one-dimensional bichromatic optical lattices~\cite{Roati} or generated electrically in quantum wires~\cite{gang12}.

\section*{Acknowledgements}

This work is supported by the NSF of China (Grants Nos.~11304280, 91221302 and 11274364) and NBRP of China (Grants Nos.~2012CB821402 and 2012CB921303).

\end{document}